\documentclass{article}

\usepackage{arxiv}

\usepackage[utf8]{inputenc} 
\usepackage[T1]{fontenc}    
\usepackage{hyperref}       
\usepackage{url}            
\usepackage{booktabs}       
\usepackage{amsfonts}       
\usepackage{nicefrac}       
\usepackage{microtype}      
\usepackage{lipsum}
\usepackage{graphicx}
\usepackage{subfigure}
\graphicspath{ {./images/} }
\usepackage{amsmath}

\title{Advance quantum image representation and compression using DCTEFRQI approach}

\author{
Md Ershadul Haque \\
  School of Coumputing, Mathematics and Engineering\\
  Charles Sturt University\\
  Bathurst, NSW 2795 \\
  \texttt{mhaque@csu.edu.au} \\
   \And
Manoranjan Paul \\
  School of Coumputing, Mathematics and Engineering\\
  Charles Sturt University\\
  Bathurst, NSW 2795 \\
  \texttt{mpaul@csu.edu.au} \\
  \And
 Anwaar Ulhaq \\
  School of Coumputing, Mathematics and Engineering\\
  Charles Sturt University\\
  Bathurst, NSW 2795 \\
  \texttt{aulhaq@csu.edu.au} \\
    \And
Tanmoy Debnath \\
  School of Coumputing, Mathematics and Engineering\\
  Charles Sturt University\\
  Bathurst, NSW 2795 \\
  \texttt{tdebnath@csu.edu.au} \\
}
\begin{document}
\maketitle
\begin{abstract}
In recent year, quantum image processing got a lot of attention in the field of image processing due to opportunity to place huge image data in quantum Hilbert space. Hilbert space or Euclidean space has infinite dimension to locate and process the image data faster. Moreover, several researches show that, the computational time of quantum process is faster than classical computer. By encoding and compressing the image in quantum domain is still challenging issue. From literature survey, we have proposed a DCTEFRQI (Direct Cosine Transform Efficient Flexible Representation of Quantum Image) algorithm to represent and compress gray image efficiently which save  computational time and minimize the complexity of preparation. The objective of this work is to represent and compress various gray image size in quantum computer using DCT(Discrete Cosine Transform) and EFRQI (Efficient Flexible Representation of Quantum Image) approach together. Quirk simulation tool is used to design corresponding  quantum image circuit. Due to limitation of qubit, total 16 numbers of qubit are used to represent the gray scale image among those 8 are used to map the coefficient values and the rest 8 are used to generate the corresponding coefficient position. Theoretical analysis and experimental result show that, proposed DCTEFRQI scheme provides better representation and compression compare to DCT-GQIR, DWT-GQIR and DWT-EFRQI  in terms of PSNR(Peak Signal to Noise Ratio) and bit rate.
\end{abstract}


\section{Introduction}\label{Intro}
In the field of quantum, computer science, physics and mathematics plays an vital roles to concrete the area of quantum information processing (QIP) \cite{b1}. In the quantum domain, Hilbert space provides plenty of space to map quantum information. In quantum system, quantum  mechanics mainly deals with the quantum properties. Entanglement and superposition are two main  properties in quantum mechanics that's provide  faster computation \cite{b2}. Now a days, due to faster computation and remarkable quantum properties, its has been gaining more research interest wordwide.  On the other hand, quantum parallelism is the inherent  phenomenon which make its unique and proven as faster compare to classical computer\cite{b3,b4}. The limitations of classical computer are given below\cite{bb6,bb8}:

\begin{itemize}
    \item Unable to solve NP hard problem rapidly.
    \item Finding the pattern work is completely routine and require no details understanding of subject of the problem.
    \item Slow computational time compared to quantum computer.
    \item Optimization – Optimization is finding out best solution to a problem among many possibilities.
\end{itemize}

According to Moore's law , the computing power of classical computer has increased in the past decade.  After that, it computing power has not increased significantly due to limitation of the several objective factor \cite{bb5}. Therefore,  it is demand of time, to increase the computing power. Feynman et al. explored the  first quantum computer other way to increase the computing power which  attained popularity in the research and  development community \cite{bb4}. In \cite{b5}, Shor proposed an algorithm for factorial calculation for integer in quantum computer which showed faster computation compare to classical computer.  After that,  following Shor algorithm, Grover provided an algorithm for database  search in quantum domain \cite{b6}. The advantages of the quantum computer are given below \cite{bb6,bb7,bb8, bb9}.

\begin{itemize}
    \item It can rapidly solve NP(non-deterministic polynomial) problem.
    \item In term of Hardware, its capable to every possible processing answer simultaneously. 
    \item Faster computational time than classical computer
    \item Can carry enormous amount of information on account of advantages of exponential formula $(2^n)$.
    \item Able to create mathematical creativity automatically.
    \item Dramatically increase the speedup in case of cryptographic codes and internet base monetary transaction
\end{itemize}
Limitations of  quantum computer\cite{bb6,bb7,bb8,bb9} are listed below:
\begin{itemize}
    \item For others application, such as chess playing, airline flight scheduling and providing theorem, the quantum computer still suffering same algorithmic limitation such as today classical computer.
    \item Generate error due to unwanted interaction between quantum computer and its environment typically know as decoherence.
    \item Unfortunately, there is catch. During measurement, its able to display only one result among lots of possibilities. All the others result will than disappear.
    \item As transistors in microchips approach the atomic scale, ideas from quantum computing are likely to become relevant for classical computing as well.  
\end{itemize}
Although there have some pros and cons, still there are lot of area of application and development of quantum computer. Below are some application area of quantum computer\cite{bb8, bb9,bb10}:
\begin{itemize}
    \item What pattern might exit in stock market?
    \item Recording of weather or brain activity. 
    \item Drug discovery process.
    \item Cryptographic systems.
    \item On store selves at the same time as wrap-drive generators and anti-gravity shields.
    \item Chemical simulation.
   \item Radar making.
   \item Weather forecasting.
\end{itemize}
Applications of quantum computer in term of image processing are recorded below \cite{bb6, bb910}:
\begin{itemize}
    \item Image compression.
    \item Image segmentation.
    \item Edge detection.
    \item Security and denoising of images.
    \item Information security.
    \item Watermarking.
    \item High privacy. 
    \item QCNN for image processing.
\end{itemize}   

In order to resolve the presentation and processing issue of classical image on quantum computer are gradually accelerating the combine application of quantum computer and classical computer \cite{bb5}. In many applications, image processing plays core characteristics of image operation \cite{b7}. In classical image processing, the number of operation required is quite high. The applications of the image area are increases day by day such as image pattern learning, image registration, image sensor data, agriculture, medicine, remote sensing \cite{b8}. Moreover, the  number of processing images become bulky which lead the complexity of the algorithm  in  classical computation. In classical computation, to process those huge image it's require more memory and hardware\cite{b8}.

Therefore, to process and store high amount of image data, it is necessary to find high performance algorithmic support computing platform \cite{b9}. To address high performance image processing computing issue, quantum is the right candidate which operate on qubit rather than binary bit \cite{b10}. The complexity of quantum computer for n-bit sequence is $O(n)$, where as classical computer requires $O(n*2^n)$. In addition, quantum image compression reduce the number of image operation to prepare quantum image in quantum domain.  Number of gates is the main resources in quantum system determine its complexity rather than counting qubits. Moreover, the total number of required gates determine quantum network time complexity where each need certain of time to do the operation \cite{b11}. In computing power, storage and communications, continuing cost improvements  are making more and more systems practical and compression frequently included there in order to provide cost-effective solutions.

In this work, a comparative study has done  for quantum image representation and compression through classical preparation method. In preparation, DWT(Discrete Wavelet Transform) and DCT are used to prepare the image before presenting in quantum system to investigate each capability to preparation. Than, quantize the coefficient values to make it feasible to represent and compress in quantum circuit. After that, quantized coefficient is represent in the quantum computer using proposed algorithm. Before presenting in quantum computer, quantized coefficient and its corresponding position is prepared to make it appropriate for quantum representation and compression. The contributions of this work are given below:
\begin{itemize}
    \item Various size of image able to represent and compress in quantum computer.
    \item Preparation complexity is removed due to apply pre-preparation approach. 
    \item Quantum bit rate can be calculated for any kind of images. 
    \item Image can be retrieved from its quantum circuit including both  pixel and position. 
    \item Any kind of quantum operation can be included through our approach. 
    \item Things are happening in quantum computer but resources calculation is done through classical computer.
    \item Combine application able to bring the real life quantum image application. 
\end{itemize}

The rest of the article is organized as follows. The literature survey is discussed in Section \ref{L_R}; proposed methodology is presented in section \ref{P_M}; result and its discussion is given in section \ref{R_D}. The conclusion of this work is outline in section \ref{CC}. 

\section{Literature and its summary}\label{L_R}
Quantum computing is a combine application of the quantum mechanics, computer science and mathematics  and provides a new para-diagram to the probable solution of Moore's law failure issue to increase the computational power\cite{bb5}. In quantum image processing, image presentation,  and image retrieval  are the main concern. Qubit latice is the first quantum approach that demonstrate the multi-particle quantum system for image storing and retrieving \cite{b10}. Latorre et al. proposed Real Ket that algorithm to represent quantum image\cite{b12}. 

\begin{figure}[htbp]
\centerline{\includegraphics[width=0.7\linewidth]{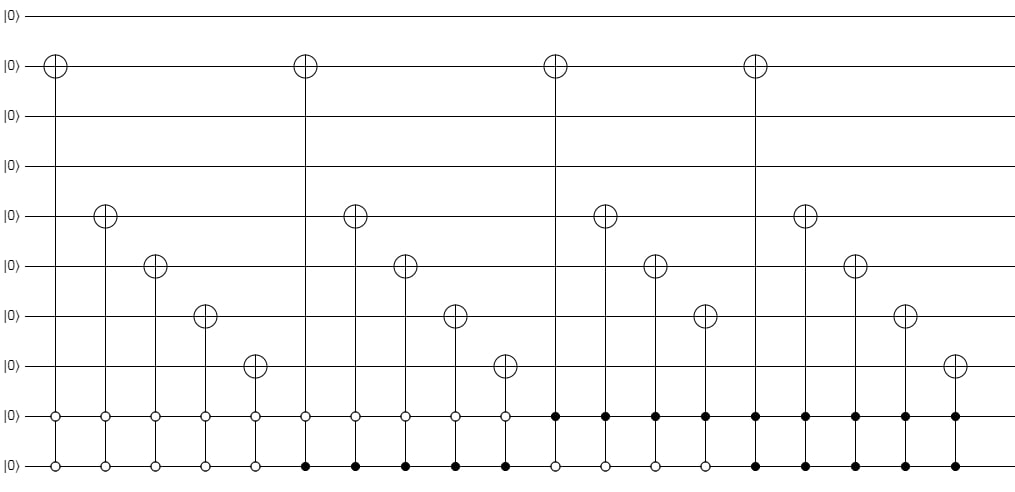}}
\caption{Quantum circuit for the GQIR representation of a $2X2$ gray scale image}
\label{fig_GQIR}
\end{figure}

Inspired with pixel reorientation of image in classical computer FRQI (Flexible Representation of Quantum Image) was developed in \cite{b13}. FRQI represent the gray scale image information as a angle in the quantum system. FRQI unable to represent pixel-wise gray scale complex operation since it used only one qubit. The Entanglement image representation was proposed in 2010 \cite{b14} to take the advantages of entanglement use. Based on JPEG(Joint Photography Expert Group) image, Jiang et al. proposed an efficient quantum compression method using GQIR  approach that use DCT preparation method in 2017 \cite{pm}. It uses DCT and GQIR approach together to increase the compression. Laurel et al. proposed quantum based equivalence pixel image from bit pixel image \cite{b15}. The NEQR (Novel Quantum Representation of Color Digital Images), was proposed  to represent the classical color image in quantum system \cite{b16}. NEQR method resolve the FRQI issue because its provide several qubit to represent gray scale image.  The limitation of NEQR is that its only able to represent square image. To solve this problem INEQR approach was proposed that able to represent unequal horizontal and vertical length image\cite{b9}. In INEQR, how color and big size image is  represent is still unclear. \\
    GQIR (General Quantum Image Representation) use logarithmic scale to represent rectangular image of arbitrary shape but its generate lot of redundant bits for position preparation\cite{b18}. For example, Table \ref{exam_GQIR} shows an $2X2$ image and Fig \ref{fig_GQIR} exhibit its corresponding $2X2$ GQRI  quantum circuit representation which directly convert pixel values and its positions into quantum circuit. To generate each pixel position location, every time similar amount of bits are required to connect each cnot gate for preparing pixel value with its state position are main drawback of this method. If more cnot gates are there to create each pixel value leading the complexity of state bit preparation.\\
\begin{table}
\caption{Example of an $2X2$ image}
\begin{center}
\begin{tabular}{|l|l|}
\hline
79 & 79 \\
\hline
79 & 79 \\
\hline
\end{tabular}
\label{exam_GQIR}
\end{center}
\end{table}

\begin{figure}[htbp]
 \centerline{\includegraphics[width=0.8\linewidth]{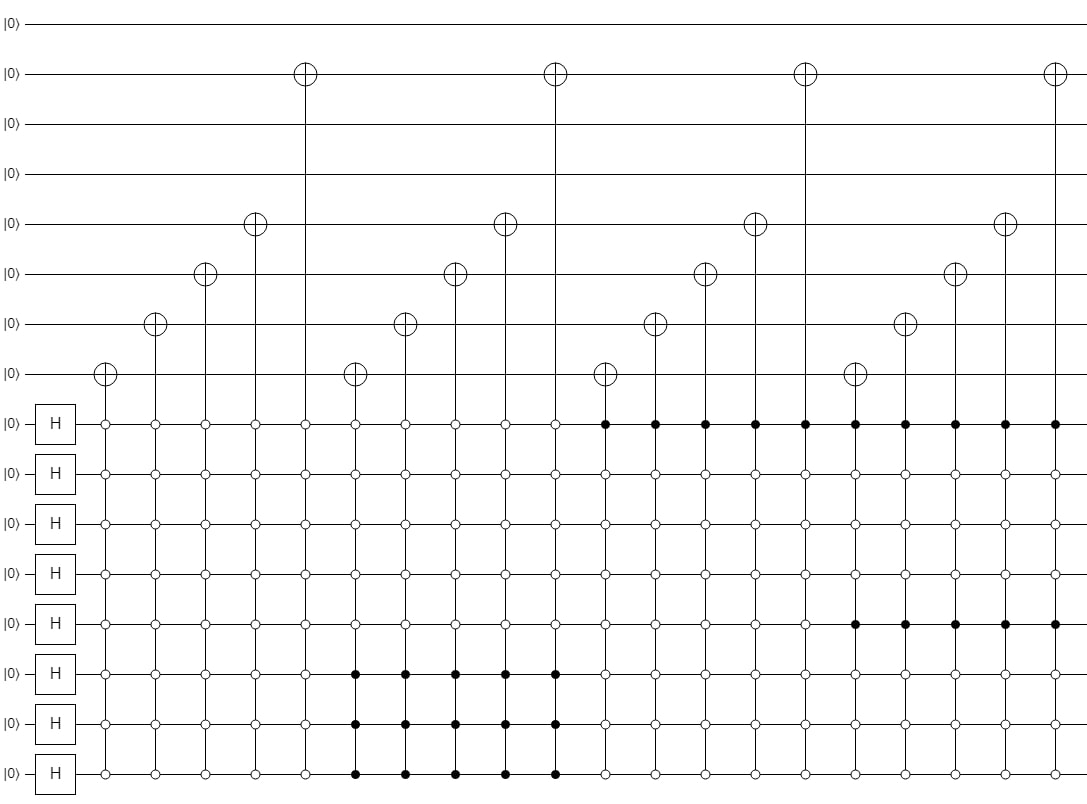}}
\caption{An 16x16 GQIR-DCT quantum image and its quantum state}
\label{fig3}
\end{figure}

In practical application, Fig. \ref{fig3} show 16x16 lena image quantum circuit as an example after performing 16  quantization using Quirk simulation tool\cite{bb8}. To prepare coefficient values in quantum domain, first convert the nonzero values of coefficient into binary stream and map the corresponding one's only using 8 numbers of qubit shown at the top of the quantum circuit in Fig.\ref{fig3}. For state preparation, corresponding position of non-zero coefficient values are also recorded and located in two dimensional (YX) position system using the rest 8 numbers of qubit shown at the bottom of Fig. \ref{fig3}. 

Polar coordinate based quantum image presentation was proposed and know as QUALPI\cite{b19}. Multidimensional represent of color image know as a NASS(normal arbitrary superposition state) was proposed \cite{b8}. In 2021, EFRQI (Efficient Flexible Representations of Quantum Image)  proposed to minimize the state preparation bits of GQIR approach\cite{b8}. However, EFRQI has the following issues:
\begin{itemize}
    \item Rather than decreasing quantum resources its increase the number of require gate compare to GQIR.
    \item provides too much complexity to represent every pixel of a medium or big size image such as $512X512$ and $1024X1024$ respectively.
\end{itemize}

\section{Proposed methodology}\label{P_M}
\begin{figure}[htbp]
\centerline{\includegraphics[width=0.7\linewidth]{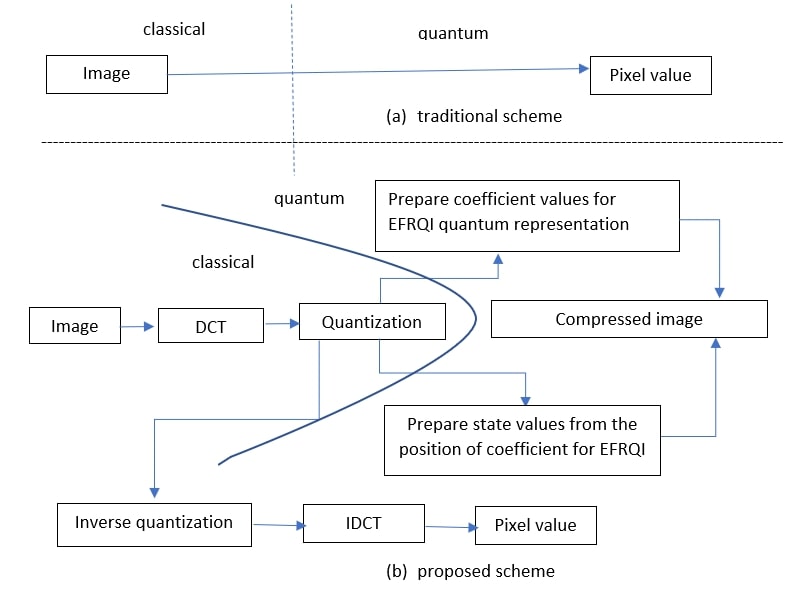}}
\caption{Basic idea of proposed scheme and difference with traditional scheme.}
\label{fig}
\end{figure}

In this section, the methodology of the proposed research is described. Fig. \ref{fig} shows the comparative view of our scheme with traditional one. In traditional scheme, pixel value is directly represent in quantum computers. In our proposed scheme, rather than  representing pixel value directly, coefficient was considered which is done via classical preparation approach. Consequently, quantized coefficient value and its corresponding position is prepared to make it suitable for quantum representation. The things are happens in quantum computer but calculation of operation is done via classical computer. In preparation step, 8 and 64 block size of DWT and 8 block size of DCT were considered including Q=8, 16, 32, 36 and 70 quantization factor to compare the performance of our proposed with traditional. 

\begin{figure}[htbp]
\centerline{\includegraphics[width=\linewidth]{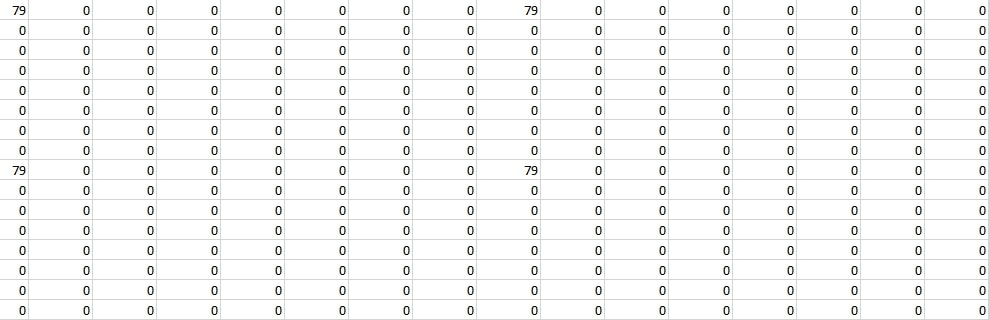}}
\caption{Example of Lena image after DCT with 16 label of quantization}
\label{fig2}
\end{figure}

\begin{figure}[htbp]
\centerline{\includegraphics[width=\linewidth]{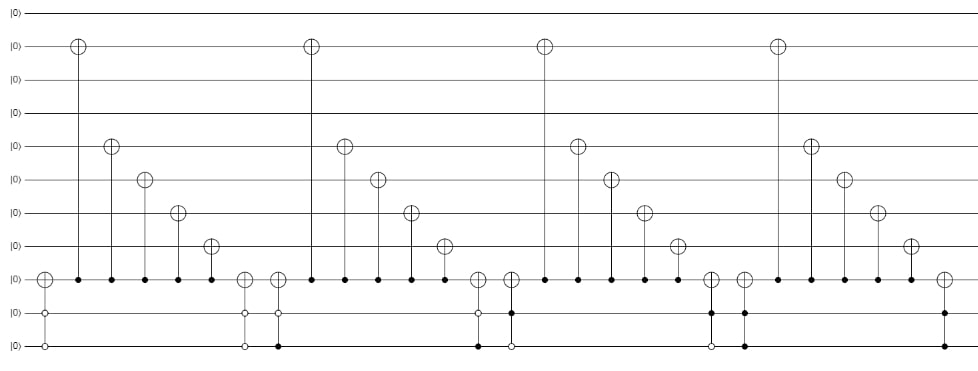}}
\caption{An 16x16 DCTEFRQI quantum image and its quantum state}
\label{PSC}
\end{figure}

Fig. \ref{fig2} shows the example of lena image after applying DCT preparation step. Hereafter, finding  non-zero  coefficient values and convert them into binary system to count how many times one's is happened in one coefficient. Since, the quantum circuit deal with one's and zero's only that governs by the qubit. The corresponding position value of each coefficient is used to prepare the quantum state. There are two kinds of bit rate; one is for coefficient and another is for its position. Fig. \ref{PSC} exhibit the corresponding quantum image of a $16X16$ lena image using proposed DCTEFRQI scheme.   

In order to decrease the complexity of preparation, compression is taken via DCT transformation rather than representing pixel value directly inside quantum computer. Block wise quantum representation of coefficient is represented in quantum computer ease the complexity of image preparation before representing in quantum system. $2^nX2^n$ image size is considered for representation and compression purpose.\\ 

In DCTEFRQI approach, steps involve are given below: \\
 
Step 1: DCT and quantize. \\
 
Step 2: Prepare the quantized DCT coefficient and make it suitable for representing in quantum circuit. This process is similar to GQIR approach. In DCTEFRQI, it require $q+2n+1$ qubits and initially set all qubits value to $\vert0\rangle$. Where, $q$ is the number of require qubits to generate non-zero coefficient value and calculated from the maximum pixel value of gray image. Moreover,  $n$ is calculated from image size for example, $n=log_2(S)$, where $S$ is the size of a square image. One auxiliary qubit act as a bridging qubit that make the connection between coefficient preparation qubits and its corresponding position representing qubits. The initial state can be explained using the below equation\cite{b8}:
 
 \begin{equation}
     \Psi_0\rangle={\vert0\rangle}^{\otimes(q+2n+1)}
 \end{equation}
 
Then, $(q+1)$ identity gates and $2n$ Hadamard gates are  used for coefficient preparation and its state preparation respectively. The identity and Hadamard matrix are shown below:
\begin{equation}
I=
\begin{pmatrix}
1 & 0 \\
0 & 1 
\end{pmatrix}
\end{equation}

\begin{equation}
H=
\begin{pmatrix}
1/\sqrt(2) & 1/\sqrt(2) \\
1/\sqrt(2) & -1/\sqrt(2) 
\end{pmatrix}
\end{equation}

In this step, the whole quantum step can be expressed as follows: 
\begin{equation}
U=I^{\otimes{q+1}}\otimes H^{\otimes{2n}}
\end{equation}

$U$ transform $\Psi_0$ from initial state to intermediate state $\psi_1$.
\begin{equation}
\Psi_1=U(|\Psi_0\rangle)=I|0\rangle^{\otimes{q+1}}\otimes H^{\otimes{2n}}
\end{equation}

The final preparation step is done using $U_2$ quantum operator:
\begin{equation}
\Psi_2=U_2(|\Psi_1\rangle)=\frac{1}{2^n} \sum_{i=1}\sum^{j=1}\,|C_{YX}\rangle |YX\rangle
\end{equation}
where $|C_{YX}\rangle$ is corresponding coefficient value and $YX$ its coordinate position. The quantum transform operator is $U_2$ is given below: 

\begin{equation}
U_2=\prod_{X=0,....,2^n-1}\prod_{Y=0,....,2^n-1}\, U_{YX}
\end{equation}

The quantum sub-operator $U_{YX}$ is also given below: 
\begin{equation}
U_{YX}= \left(I\otimes \sum_{ij\neq YX} {|ji\rangle {\langle ji|}} \right)   +\sigma_{YX} \otimes |YX\rangle {\langle YX|}
\end{equation}

The $\sigma_{YX}$ is given below:
\begin{equation}
    \sigma_{YX} =\otimes^{q-1}_{i=0}{\sigma^i_{YX}}
\end{equation}

The function of $\sigma^i_{YX}$ is setting the vallue of $i_{th}$ qubit of (YX)'s quantized DCT coefficient. 

For state preparation, Hadamard gate is used to create the superposition and c-not is used to create the entanglement between qubits in the quantum circuit. The identity gates do not have any effect on qubit initial state that mean's the original state of qubit remain unchanged. The Hadamard gate create the superposition of state $|0\rangle$  and $|1\rangle$ with equal probability. \\

 Step 3: After performing $8X8$ block DCT, store the quantized coefficient. Then, prepare the non-zero quantized coefficient for representing in the quantum  system. In the meantime, record the corresponding non-zero coefficient position for preparing its quantum state. Calculate the bit rate from  non-zero coefficient which include one's only. In addition, when calculate it's position count both frequent number of zero's  and one's happen in state connection with considering extra bit to locate the block position that minimize the block position error. In addition, sign bit also considered to assign each coefficient sign value. \\
 
 Step 4: Inverse quantization. \\
 
 Step 5: Inverse DCT. \\
 
 Step 6: Compute PSNR to qualify the reconstructed image. The PSNR is defined as follows\cite{pm}:
 
  \begin{equation}
     PSNR=20*log_{10}\frac{MAX1}{\sqrt{MSE}}
 \end{equation}
 where $MAX1$ is the maximum possible pixel value of an image. The MSE(Mean Square Error) is expressed as follows:

\begin{equation}
    MSE=\frac{1}{mn}\sum_{0}^{m-1} \sum_{0}^{n-1} ||(i,j)-g(i,j)||^2
\end{equation}

In the mean time, two times compression already happened through quantum image preparation and presentation. Firstly, compression is happened in the preparation stage and finally its happens again when its represent in quantum circuit since its consider only one's and discard all zero value to prepare the coefficient. 

\section{Result and discussion}\label{R_D}

Four types of image are shown in Fig. \ref{result_test} were selected from database sample for result verification. The details of each image are given in Table \ref{image_details}.

\begin{table}
 \caption{Selected Sample Image details}
  \centering
  \begin{tabular}{ll}
    \toprule
    Image Name          & Image Size \\
    \midrule
    Deer   & $1024X1024$    \\
    Cameraman      & $192X192$      \\
    Scenery          & $512X512$  \\
    Lena           & $512X512$  \\
    \bottomrule
  \end{tabular}
  \label{image_details}
\end{table}

\begin{figure*}[htbp]
\centerline{\includegraphics[width=\linewidth]{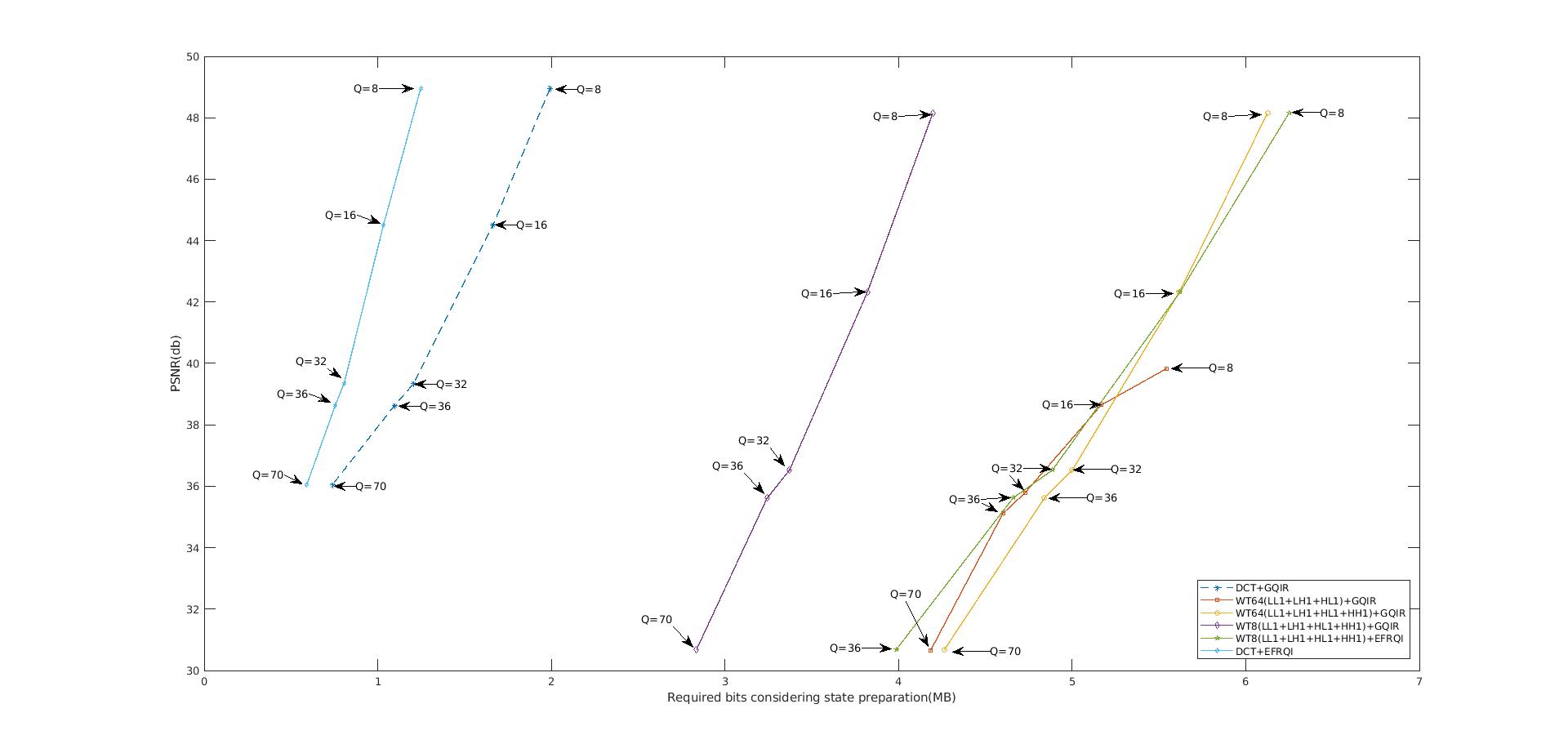}}
\caption{Bit rate versus PSNR for deer image}
\label{fig4}
\end{figure*}

Fig. \ref{fig4} shows the rate distortion curve for deer image depict in Fig.\ref{result_test}(a). The result shows for five types of different quantization factor, Q=8, 16, 32, 36, and 70. In case of DWT-GQIR, 8 block size shows the better result compare to 64 block size. In both cases, its exhibit the same PSNR but provides different bit rate although different combination of wavelet label 1 was applied. Compare to DCT-GQIR, DWT-GQIR shows the poor result over considered quantization. It happens because DWT generate lower value of  coefficient in the higher state position. On the other hand, DCT generate a higher value of coefficient in the lower position. The same thing is happened  over the whole coefficient and that's why DCT-GQIR show better result compare to DWT-GQIR. Result in Fig. \ref{fig4} show that proposed DCTEFRQI display better result compare to DCT-GQIR. Because, DCT-EFRQI able to create higher coefficient values in lower state position with the help of auxiliary qubit and Toffoli gate. Auxilary qubit was used to connect the coefficient values representing qubits to its corresponding position representing qubits via Toffoli gate. Both auxilary qubit  and Toffoli gate were contributed to lower quantum bit rate that's means lower quantum operation resources is required in DCTFRQI compare to DCT-GQIR. Therefore, DCT-GQIR and  DCTEFRQI are considered through  remaining analysis of the result that determine the performance of each combination using rest number of selected sample images. 

\begin{figure}[htbp]
\centerline{\includegraphics[width=0.6\linewidth]{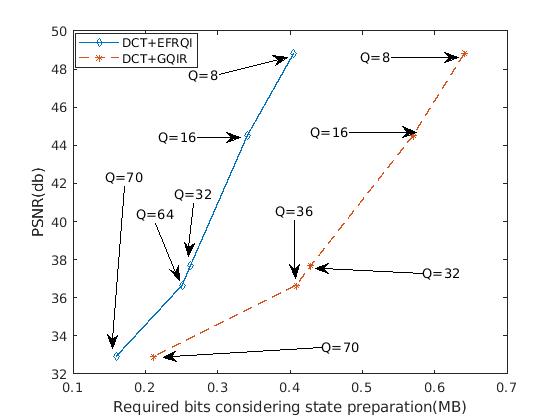}}
\caption{Bit rate versus PSNR for cameraman image}
\label{fig5}
\end{figure}

Fig. \ref{fig5} shows the result of PSNR versus bit rate of cameraman image for both DCT-GQIR and DCTEFRQI. Comparison result shows that, DCTEFRQI exhibit better result compare to DCT-GQIR over every quantization  factor, Q=8, 16, 32, 36 and 70. The bit rate make this difference while same PSNR exhibited by DCT-GQIR and DCTEFRQI. It is concluded that, DCTEFRQI obviously perform better than DCT-GQIR. 

\begin{figure}[htbp]
\centerline{\includegraphics[width=0.6\linewidth]{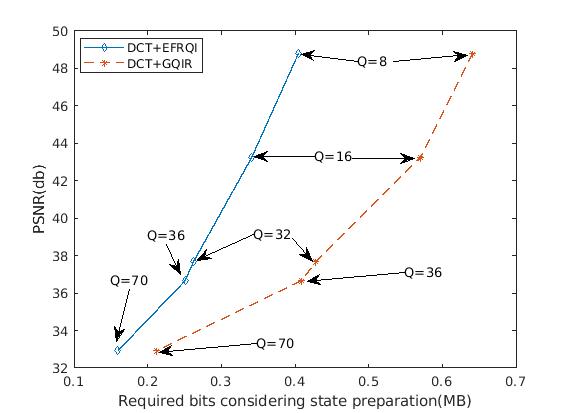}}
\caption{Bit rate versus PSNR for scenery image}
\label{fig6}
\end{figure}

Fig. \ref{fig6} shows the comparison result of PSNR versus bit rate for scenery image. Result shows that, DCTEFRQI provide better bit rate compare to DCT-GQIR but exhibit same PSNR over considered quantization factor. From this result, It is concluded that, DCTEFRQI enact efficient compression method compare to DCT-GQIR for quantization factor, Q=8, 16, 32, 36, and 70. 
 
\begin{figure}[htbp]
\centerline{\includegraphics[width=0.6\linewidth]{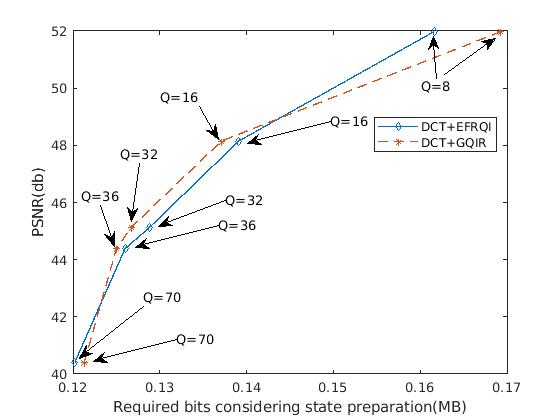}}
\caption{Bit rate versus PSNR for lena image}
\label{fig7}
\end{figure}

Fig. \ref{fig7} reveals the result of PNSR versus bit rate of lena image for both DCTEFRQI and DCT-GQIR approach. Result shows that, DCT-GQIR provide slightly better result in case of 8, 16 and 32 quantization factor in terms of bit rate. On the other hand, DCTEFRQI  act well compare to DCT-GQIR in case of 8 and 70 quantization factor. This things can be explained in such as way more information lost is happens after applying quantization factor 16, 32 and 36 in case of DCTGQIR compare to DCTEFRQI while both provide the same PSNR. From the region of improvement, it is concluded that DCTEFRQI display better result compare to DCT-GQIR approach. 

\begin{figure}[htbp]
\centerline{\includegraphics[width=0.6\linewidth]{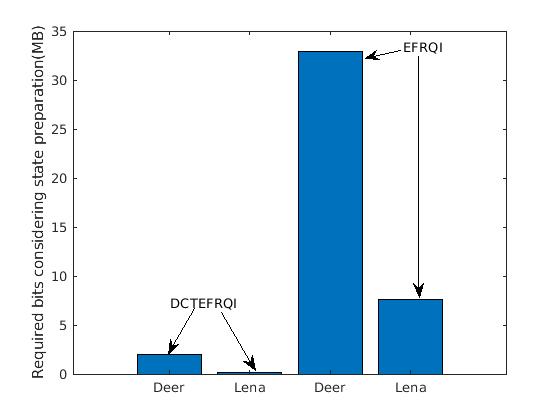}}
\caption{Bit rate comparison for deer and lena image}
\label{BitRate_Comparison}
\end{figure}

\begin{figure*}
\label{result_test}
\centering
    \subfigure [deer image]
    {
        \includegraphics[width=0.35\textwidth, height=5cm]{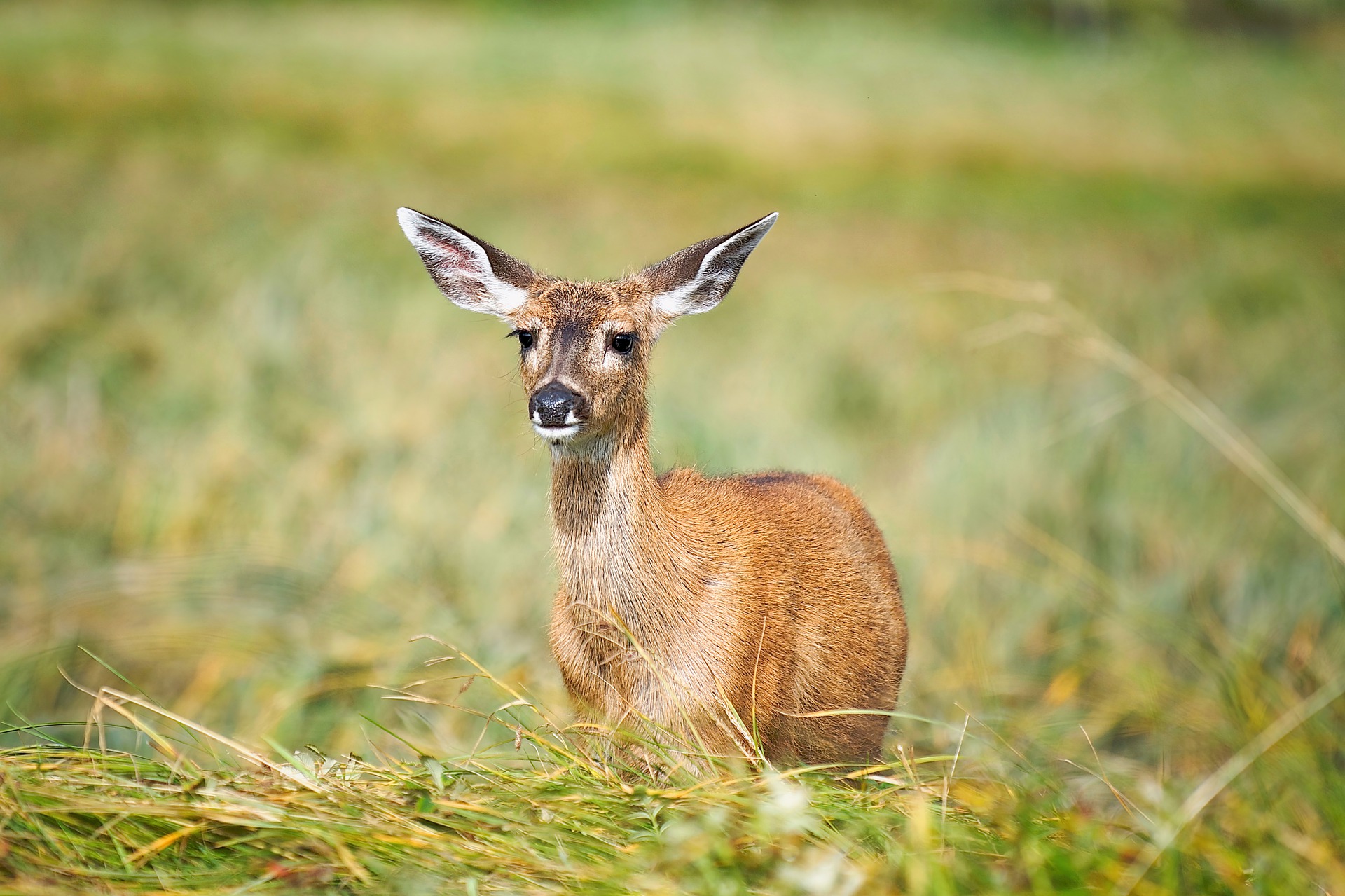}
        \label{dear}
    }
    \subfigure[cameraman image]
    {
        \includegraphics[width=0.35\textwidth, height=5cm ]{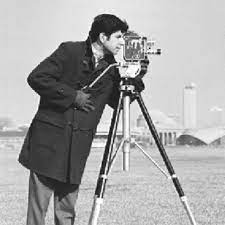}
        \label{camera}
    }\\
    \subfigure[scenery image]
    {
        \includegraphics[width=0.35\textwidth,height=5cm ]{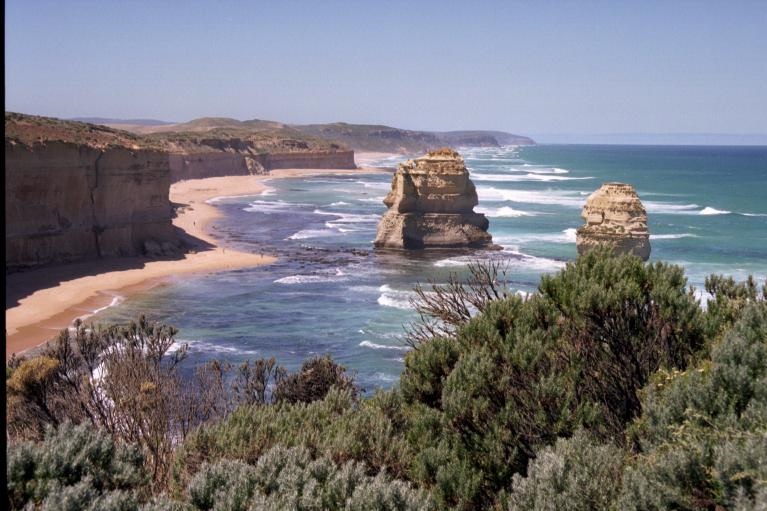}
        \label{scenery}
    }
    \subfigure[lena image]
    {
        \includegraphics[width=0.35\textwidth,height=5cm ]{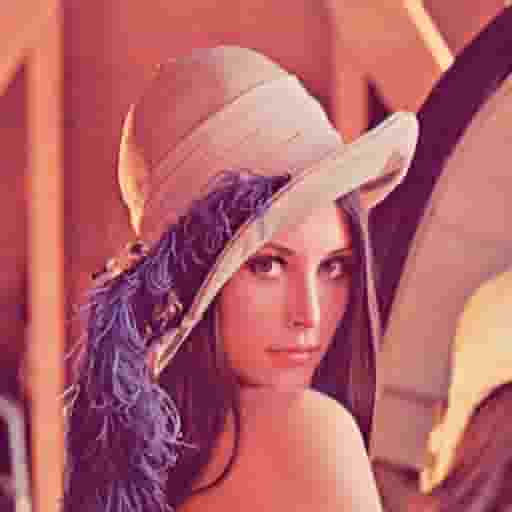}
        \label{A3_energy_Sensing_coverage}
    }
    \caption{Sample image from database \cite{bb17,bb18}.}

\end{figure*}

Fig. \ref{BitRate_Comparison} shows the bit rate comparison result of our proposed method compare to EFRQI along. Two types of sample image one is deer image and another is lena image are considered to determine the performance of bit rate. Comparison result shows that,  in case of deer image,  DCTEFRQI compress more than 16 time compare to EFRQI. In addition, in terms of lena image, DCTEFRQI able to compress more than 44 times  compare to EFRQI. From this result, it is concluded that, DCTEFRQI perform more advance for bit rate in both $512X512$ and $1024X1024$ cases of image size.   

\section{Conclusion}\label{CC}
This paper focus on the quantum image representation and compression and proposed a new quantum image compression scheme. From this work, the below things are concluded compare to the previous method \cite{pm,bb10}:
\begin{itemize}
    \item Any size of image can be represented and compressed in the quantum system.
    \item Provides better bit rate.
    \item PNSR do not has any effect if no transform or combination is happened inside quantum system.
    \item Bit rate is main fact to represent the image inside quantum computer
    \item Its simple and fast for calculating quantum operation resources. 
    \item Its open up lot of opportunity to processing and compressing of classical image inside  quantum processor.
    \item For medium size image, DCTEFRQI compress more than two times compare to high size image. 
\end{itemize}

\section*{Acknowledgment}
The author declare that there is no conflict of interest.

\bibliographystyle{unsrt}  


\end{document}